\documentclass{interact}

\usepackage{graphicx,epstopdf}
\usepackage{bm}
\usepackage{color}
\usepackage{amsmath}
\usepackage{mathtools}
\usepackage[ngerman,english]{babel}
\usepackage{amsfonts}
\usepackage{amsmath}
\usepackage{braket}

\usepackage[numbers,sort&compress,merge]{natbib}
\bibpunct[, ]{[}{]}{,}{n}{,}{,}

\newcommand{\Fe}{[Fe(H$_2$O)$_{6}$]$^{2+}$}

\newcommand{\Sch}{Schr\"{o}dinger }

\begin{document}


\articletype{FULL PAPER}

\title{Density matrix based time-dependent configuration interaction approach to ultrafast spin-flip dynamics}

\author{
\name{Huihui Wang\textsuperscript{a}, Sergey~I. Bokarev\textsuperscript{a}\thanks{CONTACT S.~I. Bokarev Email: sergey.bokarev@uni-rostock.de}, Saadullah G. Aziz\textsuperscript{b}, and Oliver K\"{u}hn\textsuperscript{a}}
\affil{\textsuperscript{a}Institut f\"{u}r Physik, Universit\"{a}t Rostock,
  Albert-Einstein-Str. 23-24, 18059 Rostock, Germany; 
  \textsuperscript{b}Chemistry Department, Faculty of Science, King Abdulaziz University, 21589
  Jeddah, Saudi Arabia}
}

\maketitle

\begin{abstract}
Recent developments in attosecond spectroscopy yield access to the correlated motion of electrons on their intrinsic  time scales.  Spin-flip dynamics is usually considered in the context of valence electronic states,  where spin-orbit coupling is weak and processes related to the  electron spin are usually driven by nuclear motion.  However, for core-excited states, where the core hole has a nonzero angular momentum, spin-orbit coupling is strong enough to drive spin-flips on a much shorter time scale. Using density matrix based time-dependent restricted active space configuration interaction including spin-orbit coupling, we address an unprecedentedly short spin-crossover for the example of L-edge (2p$\rightarrow$3d) excited states of a prototypical Fe(II) complex. This process occurs on a time scale, which is faster than that of   Auger decay ($\sim$4\,fs) treated here explicitly. Modest variations of carrier frequency and pulse duration can lead to substantial changes in the spin-state yield, suggesting its  control by soft X-ray light.
\end{abstract}


\section{Introduction}

The rapid development of   high harmonic generation techniques~\cite{Kling2008,Ivanov2009,Lepine2013} has recently enabled studies of  processes occurring in atoms, molecules, and nanoparticles, which are triggered by high-energy radiation including soft X-ray light~\cite{Dromey2006,Zhang2015,Popmintchev2015} on ultrashort time scales approaching few tens of attoseconds~\cite{Teichmann2016}.  The new light sources allow to trigger and control ultrafast electronic processes via preparation of complex superpositions of quantum states and to analyze their subsequent evolution.  Attosecond spectroscopy has a huge potential to study atomic and molecular responses to incident light~\cite{Picon_2016,Healion_2012}. It provides access to, e.g., electron correlation manifesting itself in the entanglement of bound- and photo-electrons (shake-ups), Auger and interatomic Coulomb decay, as well as to the coupling of electrons in plasmonic systems~\cite{Kling2008,Ivanov2009,Sansone2012,Lepine2013}. 

Among the most prominent examples of such processes is the oscillatory charge (hole) migration following the ionization from a localized moiety in molecules~\cite{Hennig2005,Remacle2006,Sansone2012,Kuleff2014,Nisoli2014,Kraus2015}. Remarkably, such electron dynamics is driven solely by the electron correlation, while the nuclei could be kept fixed in principle. This process can be considered as the electronic part or an elementary step of a general  charge transfer reaction~\cite{Oliver}, which necessarily involves nuclear motion required to localize the oscillating charge. One should admit, however, that predominantly theoretical studies were done using mostly time-dependent variants of configuration interaction and Green's function  approaches (for review, see Ref.~\cite{Kuleff2014}), while the experimental observation of such process is more difficult and stays scarse~\cite{Nisoli2014}. From the practical viewpoint, microscopic understanding of such ultrafast transfer phenomena  is essential, e.g., to approach the fundamental limits of the transmission speed of signals relevant for molecular electronics. 

Devices making use of the magnetic moment associated with spin are a prospective extension of conventional electronics~\cite{Dietl_Spintronics_2008,Bader2010}.  Recently, spin-crossover dynamics attracted much attention, e.g., in the context of high-density magnetic data storage devices~\cite{Cannizzo2010,Halcrow2013}.   Popular materials considered in this context are based on  Fe(II) organometallic complexes with variable ligands. Due to their partially filled 3d-shell they have  low-lying high-, intermediate- or low-spin electronic states depending on the ligand field strength. Upon valence excitation these systems exhibit an intra-atomic ultrafast spin-crossover occurring on time scales of the order of 100~fs~\cite{Auboeck2015, Sousa2013}. The spin-orbit couplings (SOC) between valence excited states, however, are small and spin-crossover is essentially driven by nuclear motion since it requires the nuclear wavepacket to pass through a region of near-degeneracy of two states of different multiplicity (for review see Ref.~\cite{Mai_2015}). Thus the time scale is determined by the related vibrational periods (see also Ref.~\cite{Bargheer_2002}). In passing we note that this is also a typical time scale for spin transfer between magnetic centers in polynuclear metal complexes~\cite{Jin_2012}.

For 2p core-excited electronic states of transition metal complexes, however, the magnitude of SOC increases dramatically.  Therefore, one expects the spin dynamics  to change from a nuclear to an electronically driven process.  It was recently demonstrated~\cite{Wang_arxiv} that electronically driven spin-crossover after core excitation in transition metals indeed takes place on a few femtosecond time scale, thus, being faster than the core-hole lifetime. This process can be considered as an elementary step of the conventional nuclear dynamics driven spin-crossover~\cite{Cannizzo2010}, analogously to the above mentioned charge migration. In both cases, electronic wave packet dynamics is ultimately coupled to nuclear motions, eventually leading to charge or spin localization. 

Here, we further elaborate on this finding, by reformulating the time-dependent restricted active space configuration interaction method with account for SOC within the open system density matrix formalism. Inclusion of strong correlation effects into the model is essential at this point, since transition metal complexes are known to have a multi-configurational nature of the wave function, sometimes even in the ground state. The importance of strong SOC and multi-reference effects especially applies to the 2p core-excited electronic states. In the present work, we study the influence of different excitation regimes and pulse characteristics as well as phenomenological inclusion of Auger decay. The article is organized as follows: First, we present the theoretical model and computational details in Sections \ref{sec:theory} and \ref{sec:comp}. Subsequently, we discuss the results of the application of the developed methodology to the prototypical Fe(II) compound \Fe representing a model for the solvated Fe$^{2+}$ ion in Section \ref{sec:results}. Conclusions and outlook are given in Section \ref{sec:conclusions}
%
\section{Theory}
\label{sec:theory}
Ultrafast spin-flip is investigated using the Time-Dependent Restricted Active Space Configuration Interaction method in its  density matrix formulation ($\rho$-TD-RASCI), which is similar in spirit to the techniques proposed in Refs.~\cite{Tremblay2011,Kato2012,Jin_2012}. As compared to TD-RASCI, the density matrix formulation offers a convenient way of treating dissipative dynamics of open quantum systems.
Working within Born-Oppenheimer approximation and provided that processes under study occur much faster than the period of nuclear motion we assume the clamped nuclei approximation. Here, the nuclei are fixed at the ground state equilibrium positions, and thus we solve the electronic \Sch equation only. The question whether the nuclear motion can be neglected for the early-time dynamics has triggered an ongoing debate~\cite{Mendive-tapia_2013,Li2015, Vacher_2015, Despre_2015}. Here, we assume that the system is excited far from conical intersections and that the considered time interval is shorter than the relevant vibrational periods.

In the present case of the interaction of a molecular system with X-ray light, where electronic transitions possess  a very local character, the focus is put on a fairly small subsystem containing the absorbing atom with its first coordination shell.
To account for dissipation due to the more extended environment or electronic relaxation processes, which are not treated explicitly, it is natural to represent the system in terms of the reduced density operator ($\hat\rho$) evolving in time according to

 \begin{equation}\label{eq:LvN}
    \frac{\partial}{\partial{t}}\hat{\rho}=-i[\hat{H},\hat{\rho}]+\mathcal{D}\hat\rho\,.
  \end{equation}
Here, $\hat{H}$ is the Hamiltonian operator of the subsystem of interest and $\mathcal{D}$ is the dissipation superoperator, which accounts for different dissipation processes.

The  reduced density operator is represented in the basis of Configuration State Functions (CSFs), $\ket{\Phi_j^{(S,M_S)}}$, with the total spin $S$ and its projection $M_S$:
%
\begin{equation}\label{eq:rho}
\pmb{\rho} (t) =\sum_{j,j'}{\rho_{j,j'}^{(S,M_S,S',M_S')}(t) \ket{\Phi_j^{(S,M_S)}}\bra{\Phi_{j'}^{(S',M_S')}}} \, .
\end{equation}
%
The CSFs are constructed using a time-independent molecular orbital (MO) basis, optimized at the restricted active space self-consistent field~\cite{Malmqvist1990} level, prior to propagation. 
The Hamiltonian in the CSF basis reads
%
\begin{eqnarray}
	\label{eq:Ham}
   \mathbf{H}(t)&= &\mathbf{H}_{\rm CI}+\mathbf{V}_{\rm SOC}+\mathbf{U}_{\rm ext}(t) \nonumber \\
   &=&
   \left(
   \begin{array}{cc} 
      \mathbf{H}_{h}  &  0 \\
      0  & \mathbf{H}_{l} \\
   \end{array}
   \right)
   +   
   \left(
   \begin{array}{cc} 
      \mathbf{V}_{hh}  &  \mathbf{V}_{hl} \\
      \mathbf{V}_{lh}  &  \mathbf{V}_{ll} \\
   \end{array}
   \right) 
   +
  \left(
  \begin{array}{cc} 
      \mathbf{U}_{h}(t) &  0 \\
      0  &  \mathbf{U}_{l}(t) \\
   \end{array}
   \right),
\end{eqnarray}
where we separated blocks of low ($l$) and high ($h$) spin states. 
In Eq.~\eqref{eq:Ham}, $\mathbf{H}_{\rm CI}$ is the configuration interaction (CI) Hamiltonian containing the effect of electron correlation. The eigenstates of this spin-free Hamiltonian $\mathbf{H}_{\rm CI}$ will be called spin-free (SF) states. 
The structure of the SF state can be viewed as linear combination
\begin{equation}\label{wave function}
    |\Psi^{\rm SF}\rangle=C_0 \ket{\Psi_{0}}+\underbrace{\sum_{ia}C_{i}^{a}\ket{\Psi_{i}^{a}}}_{\rm 1h1p}+\underbrace{\sum_{i<j,a<b}C_{ij}^{ab}\ket{\Psi_{ij}^{ab}}}_{\rm 2h2p}+\ldots,
\end{equation}
where $\ket{\Psi_{0}}$ represents the reference wave function of the ground electronic state and the remaining terms are the single (1h1p), double (2h2p), etc excitations from the ground state configuration. 
Indices $i,j,\ldots$ denote occupied MOs from which electron is excited to unoccupied $a,b,\ldots$ ones.
Employing the restricted active space strategy one can conveniently choose the basis of CSFs involving excitations from/to orbitals of interest, being systematically improvable up to the exact limit. 

SOC is contained in $\mathbf{V}_{\rm SOC}$, whose matrix elements are calculated within a perturbative LS-coupling scheme making use of atomic mean-field integral approximation~\cite{AMFI1996}. The latter is an effective one-electron approximation to the Breit-Pauli equation, which has demonstrated good performance for  L-edge spectra of transition metal compounds~\cite{Josefsson2012,Bokarev2013,Suljoti_2013,Atak2013,pinjari_2014,engel_2014,Bokarev2015,wernet_2015,Grell2015,pinjari_2016,Golnak2016,Preuse_2016}.  
It   provides an intuitive interpretation in terms of pure spin-states with well-defined $S$ and $M_S$ quantum numbers. The eigenstates of $\mathbf{H}_{\rm CI}+\mathbf{V}_{\rm SOC}$ are called SOC states and are linear combinations of the SFs:
 \begin{equation}\label{eq:SOC_states}
 \ket{\Psi^{\rm SOC}_{k}}=\sum_{n}a_{kn}^{(S,M_S)} \ket{\Psi^{{\rm SF},(S,M_S)}_{n}} \, .
\end{equation}

The interaction with the time-dependent electric field, 
\begin{equation}
\mathbf{U}_{i}=-\vec{\mathbf{d}}_{ii} \cdot \vec{E}(t) \, ,
\end{equation}
is taken in  semi-classical dipole approximation with the transition dipole matrices $\vec{\mathbf{d}}_{ii}$. The field vector, $\vec{E}(t)$, has polarisation $\vec{e}$, carrier frequency $\Omega$ and Gaussian envelope: 
\begin{equation}\label{eq:pulse} 
\vec{E}(t)=\vec{e}E_0\cos(\Omega t)\exp(-t^2/(2\sigma^2)).
\end{equation} 

The description of  L-edge states of transition metals requires to take into account the Auger autoionization channel, which  is known to dominate the decay process~\cite{stoehr_nexafs_1992}. This is  incorporated  at the level of phenomenological decay rates similar in spirit to Ref.~\cite{Tremblay2011}. It is convenient to define the decay constants, $\Gamma_k$, in the basis of eigenstates $\ket{\Psi^{\rm SOC}_k}$ of $\mathbf{H}_{\rm CI}+\mathbf{V}_{\rm SOC}$ yielding the simple dissipation operator:
\begin{equation}\label{eq:auger}
\mathcal{D^{\rm SOC}}=-\sum_k \Gamma_k \ket{\Psi^{\rm SOC}_k} \bra{\Psi^{\rm SOC}_k}.
\end{equation}
The respective $\mathcal{D^{\rm SOC}}$ matrix is then transformed to the CSF basis and used for the propagation (Eq.~\eqref{eq:LvN}).

\section{Computational details}
\label{sec:comp}

\begin{figure}
\centering
\includegraphics{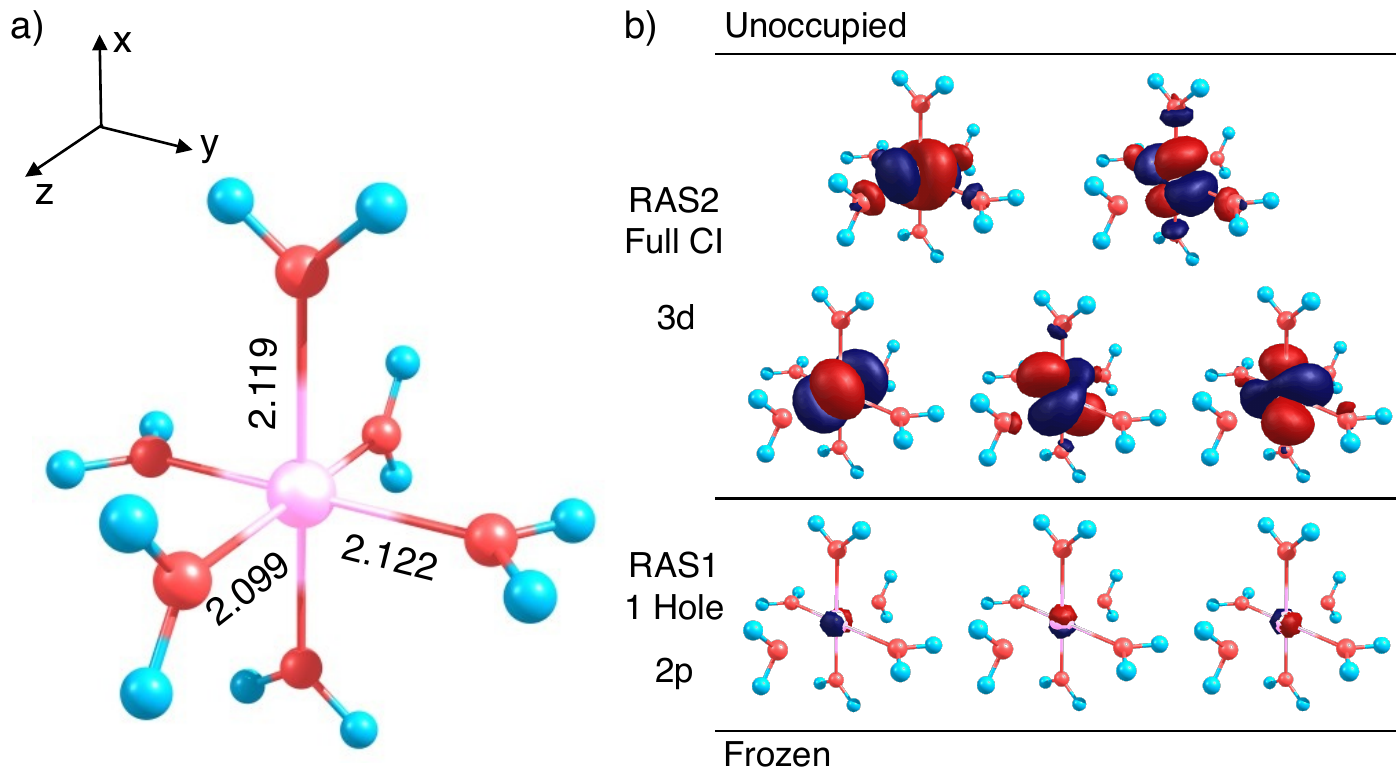}
\caption{a) General view of the \Fe complex; the lengths of three pairs of Fe--O bonds are also given (in \AA). b) MO active space used for the TD-RASCI calculation.
\label{fig:system}
}
\end{figure}

The outlined approach is applied to the spin-dynamics in the \Fe complex (see Fig.~\ref{fig:system}a)) representing a model of the solvated Fe$^{2+}$ ion, whose X-ray absorption and resonant inelastic X-ray scattering spectra were discussed in Refs.~\cite{Bokarev2013,Atak2013,Golnak2016}.
The dynamics, first described in Ref.~\cite{Wang_arxiv} and discussed below, is driven solely by the electronic coupling (SOC and electron correlation), while nuclei are fixed at the ground state equilibrium positions taken from Ref.~\cite{pierloot_relative_2006}. To justify this approximation, we note that the considered time interval of 15~fs  is shorter than the relevant vibrational periods. For \Fe, the  Fe--O stretching and O--Fe--O deformation modes, potentially influencing the  2p$\rightarrow$3d core excited electronic states, have periods above~100~fs.

The active space used in the $\rho$-TD-RASCI calculations contained 12 electrons distributed over the three 2p (RAS1 subspace) and five 3d (RAS2 subspace) orbitals (cf. Fig.~\ref{fig:system}b)) to describe the core excited electronic states corresponding to the dipole allowed 2p$\rightarrow$3d transitions~\cite{Bokarev2013,Atak2013,Grell2015,Golnak2016}. The number of holes in RAS1 was limited to one, whereas full CI was done within RAS2.  This active space included up to 4h4p configurations and resulted in 35 quintet ($S=2$) and 195 triplet ($S=1$) electronic states, directly interacting via SOC according to the $\Delta S=0,\pm1$ selection rule.  Note that septet electronic states ($S=3$) are not possible with this active space. Accounting for the different $M_S$ components, the total amount of the SF and SOC  states was 760, where 160 are valence and 600 core ones. The respective calculations are denoted as RASCI(1,2) below. Notice that both, the account for  4h4p excitations and SOC are essential to recover the  dynamics of the highly correlated  core-excited states. In addition singlet states were also included  to test the influence of second-order SOC effects; denoted as RASCI(0,1,2). Their number increased the dimensionality of the basis by 170 states in total (50 valence and 120 core).  To account for scalar relativistic effects, the Douglas-Kroll-Hess transformation~\cite{Douglas_1974} up to second order was utilized. To correct for weak correlation effects the single-state second-order perturbation theory correction (RASPT2)~\cite{Malmqvist_2008} was  added to the diagonal of $\mathbf{H}_{\rm CI}$ matrix written in the SF basis, which was then back-transformed to the CSF basis. Thus, due to their diagonal nature this corrections influenced SOC only implicitly via the relative energetics of the interacting states. To avoid intruder states in RASPT2 calculations, an imaginary level shift~\cite{Forsberg_1997} of 0.4 E$_h$ was introduced. 1s,2s, 3s, and 3p orbitals of iron as well as 1s orbitals of oxygen were kept frozen in RASPT2.

Evaluation of the  $\mathbf{H}_i$, $\mathbf{V}_{ij}$, and $\vec{\mathbf{d}}_{ii}$ matrix elements in Eq.~\eqref{eq:Ham} was performed with a locally modified version of the MOLCAS 8.0~\cite{Aquilante2016} quantum chemistry suite, applying the relativistic ANO-RCC-TZVP basis set~\cite{Basis1,Basis2} for all atoms.
Phenomenological Auger decay rates $\Gamma_k$ were set to 0.4 and 1.04\,eV for the L$_3$ and L$_2$ edges, respectively~\cite{Ohno_2009}. These decay rates correspond to lifetimes of the core hole of 10.3 and 3.98\,fs. The choice of the excitation regimes and pulse parameters in Eq.~\eqref{eq:pulse} is discussed in Section \ref{sec:results}.
The reduced density matrix equation,  Eq.~\eqref{eq:LvN},  was solved using the Runge-Kutta-Fehlberg integrator, thus utilizing adaptive step size control.  The time step varied from 2.5\,as, when the external field was small or absent, down to 0.09\,as when the field strength was substantial.

\section{Results}
\label{sec:results}

\begin{figure}
\centering
\includegraphics[width=1.0\textwidth]{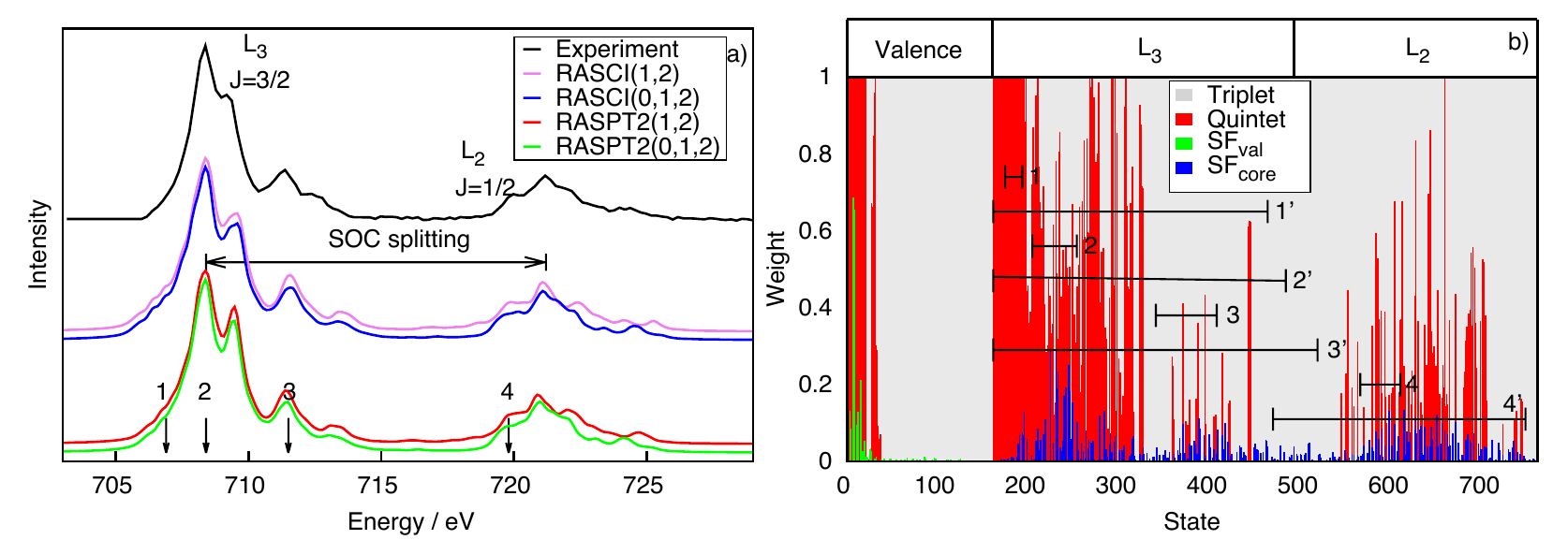}
\caption{a) X-ray absorption spectrum of \Fe as calculated with different methods (see text) and compared with experiment (partial electron yield from 2p3d3d channel)~\cite{Golnak2016}. Arrows denote the excitation energies considered in regime II, see text. 
b) Decomposition of the SOC eigenstates into  quintet ($\sum_{n,M_S}{|a_{mn}^{(S=2,M_S)}|^2}$, red bars) and 
triplet ($\sum_{n,M_S}{|a_{mn}^{(S=1,M_S)}|^2}$, grey bars) SF states (cf. Eq.~\eqref{eq:SOC_states}). The particular contributions of  valence SF (SF$_{\rm val}$, green bars)  and core  SF  (SF$_{\rm core}$ blue bars) states used in regime I to the different SOC states are also shown.  Numbered ranges reflect the bandwidths of 0.5\,eV (1-4) and 5.0\,eV (1'-4') pulses with carrier frequencies denoted in panel a) in terms of the involved SOC states. 
\label{fig:xas}
}
\end{figure}

In Fig.~\ref{fig:xas}a) the L-edge absorption spectrum of \Fe as calculated with different methods is compared to experiment~\cite{Golnak2016}.  It has a shape characteristic for transition metals, featuring  the L$_3$ ($J=3/2$) and L$_2$ ($J=1/2$) bands split due to the  SOC. This splitting is 12.7~eV (SOC constant is 8.5\,eV) what corresponds to a timescale of about 0.33~fs. Here RASCI(1,2) and RASPT2(1,2) denote the results of respective methods taking triplets ($S=1$) and quintets ($S=2$) states into account. They are compared to the results including singlet states ($S=0$). One observes a fairly good agreement of the shape of the spectrum, with different theoretical methods giving very similar results. The influence of the perturbation theory correction and inclusion of singlets on the spin dynamics will be discussed below.

Panel b) of Fig.~\ref{fig:xas} illustrates the degree of spin-mixing for the SOC states (Eq.~\eqref{eq:SOC_states}) in the case of RASCI(1,2), which is the main setup discussed in the following. It can be seen that the valence excited states are mostly pure quintets or triplets. In contrast the core excited states are dominantly spin mixtures, with the degree of spin-mixing varying with state energy. The low-energy flank of the L$_3$-band contains pure quintets, whereas moving to higher energies the contributions of triplets start to dominate. In the present study, we assume ultrafast preparation of the superposition of strongly spin-mixed 2p core hole states in different regimes. Thus, Fig.~\ref{fig:xas}b) will provide  a reference for the spin-nature of the involved eigenstates.

Below we discuss the three different excitation regimes, illustrating the dynamics of ultrafast spin-crossover. In \textit{regime I}, $\vec{E}(t)=0$ and it is assumed that a particular SF state, which is a superposition of strongly spin-mixed 2p core hole states~\cite{Bokarev2013,Bokarev2015}, has been instantaneously prepared. This somewhat artificial initial condition will serve as a reference, which highlights the spin dynamics driven solely by SOC. 

In \textit{regime II}, the system is initially in the ground state with the $M_S$-components of the lowest closely lying electronic states being populated according to the Boltzmann distribution at 300~K: 
\begin{equation}\label{eq:rho0}
\pmb{\rho}_0=\mathrm{diag} \{Z^{-1}\exp (E_i/kT)\},
\end{equation}
where $Z=\sum_i \exp (E_i/kT)$. In fact, the three lowest quintet states have non-negligible population, which yields  15 $M_S$ microstates. The core hole is created and thus spin dynamics is driven by an ultrashort X-ray pulse $\vec{E}(t)$  linearly polarized along different directions.

Finally, in  \textit{regime III} an instantaneous population of the particular CSF is addressed, being a strong superposition of the SF eigenstates, while the SOC and external field are switched off ($\mathbf{V}_{\rm SOC}=\mathbf{0}$ and $\vec{E}(t)=0$).
This regime corresponds to a sudden excitation and allows to explore the effect of electron correlation separately from SOC. It is similar to those regarded before for hole migration, for a review see Ref.~\cite{Kuleff2014}.

\subsection{Regime I}
\label{sec:regimeI}

We have chosen two representative quintet SF states, i.e.\ number 7 and 111, as initial states for investigating the SOC-driven spin dynamics which are denoted as SF$_{\rm val}$ and SF$_{\rm core}$ below. For the contributions of SF$_{\rm val}$ and SF$_{\rm core}$ to the stationary SOC eigenstates see  Fig.~\ref{fig:xas}b). Other states demonstrated very similar dynamics and are not considered further.  

\begin{figure}
\centering
\includegraphics{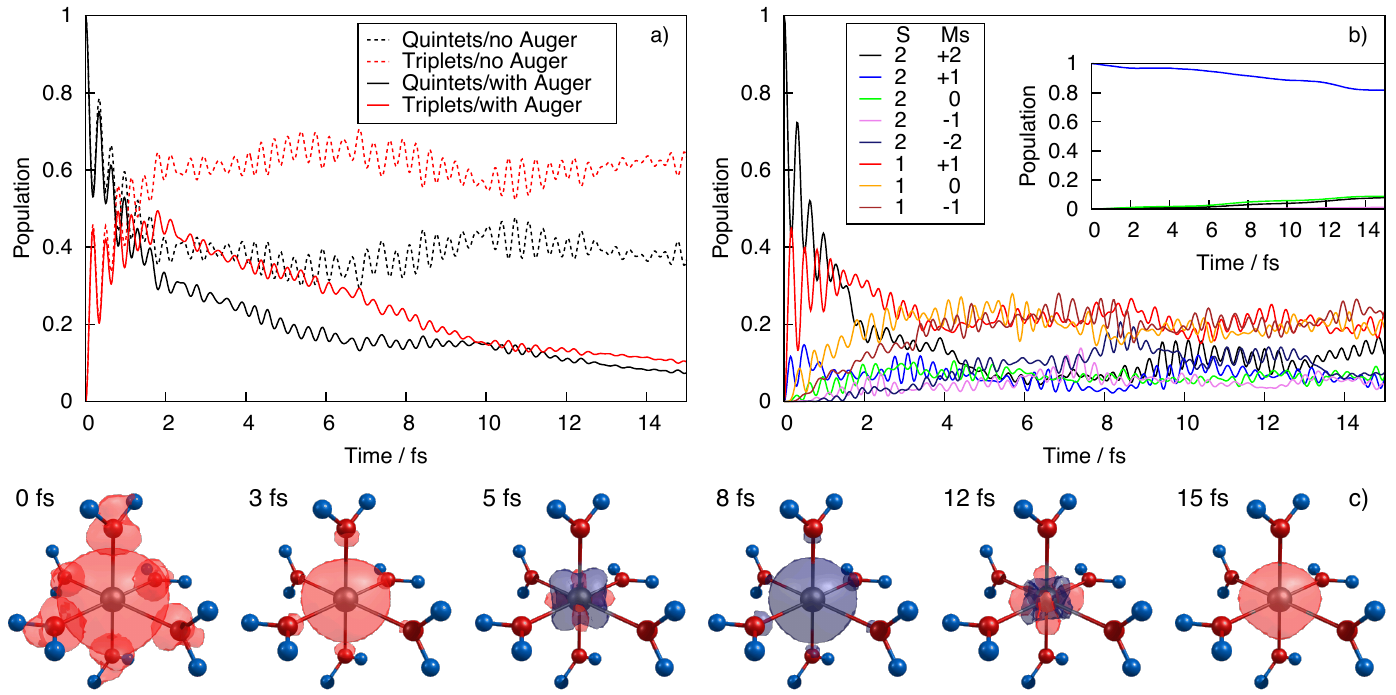}
\caption{a) Evolution of the total population of the quintet ($S=2$) and triplet ($S=1$) electronic states after instantaneous excitation to the SF$_{\rm core}$ state (regime I). Results with and without Auger decay are presented. 
b) Same as a) without Auger decay for the total population of the $M_S$ components of quintet  and triplet electronic states; initial state SF$_{\rm core}$ has $M_S=+2$. The inset gives the analogous evolution after instantaneous excitation to the valence-excited quintet  SF$_{\rm val}$ state ($M_S=+1$), demonstrating much slower dynamics.
c) Evolution of spin-density difference, $\rho_{\uparrow}-\rho_{\downarrow}$ (red - positive, blue - negative, contour value 0.001) for the case shown in panels a) and b).
\label{fig:regimeI}
}
\end{figure}

Preparation of SF$_{\rm core}$, which corresponds to   $M_S=+2$ (four spin-up electrons) and has contributions of SOC states from essentially the whole core hole excited L$_3$ and L$_2$ bands (see blue bars in Fig.~\ref{fig:xas}b), demonstrates intricate dynamics. 
It is illustrated in Fig.~\ref{fig:regimeI}, where the total populations of all quintet and triplet states with (solid lines) and without (dashed lines) Auger decay  (panel a) as well as the  detailed evolution of different $M_S$ components (panel b) are shown. As a consequence of strong SOC, the  population spreads over both quintet and triplet states such that the total triplet population becomes even larger than the corresponding quintet one within about 1~fs (Fig.~\ref{fig:regimeI}a)). The  population transfer occurs according to the $\Delta M_S=0,\pm1$ selection rule within and between both spin manifolds. The main contribution to the fast drop of the quintet population during first few fs is due to the $(S=2,M_S=+2)\rightarrow(S=1,M_S=+1)$ transitions (cf.~Fig.~\ref{fig:regimeI}b)). Quintets with $M_S=-1$ and $-2$ start to be populated only after about 1\,fs. 

The Auger decay results in a biexponential damping of the populations (Fig.~\ref{fig:regimeI}a)) resembling the choice of only two decay parameters, $\Gamma_k$, in Eq.~\eqref{eq:auger} common for all L$_3$ and L$_2$ states. Interestingly, the spin-flip occurs faster than the fastest Auger decay of 4\,fs.

Panel c) of Fig.~\ref{fig:regimeI} shows snapshots of the spin-density difference, $\rho_{\uparrow}-\rho_{\downarrow}$, evolution.  Because of this quintet-triplet population transfer, $\rho_{\uparrow}$ notably decreases during the first 3~fs.  After about 4\,fs the system almost equilibrates, i.e.\ the 760 electronic states act like an ``electronic bath'', and the corresponding populations of $M_S$ microstates oscillate around their mean value (Fig.~\ref{fig:regimeI}b)).
The spin density changes relatively slowly  from the dominating $\rho_{\uparrow}$ to the dominating $\rho_{\downarrow}$ and back due to the partial revivals of the quintet's positive and negative spin projections.
The fast modulation in Figs.~\ref{fig:regimeI}a) and b) with a period of $\approx$0.32~fs can be assigned to the SOC splitting between the L$_2$ and L$_3$ bands. It is roughly the same for all interacting states and is an intrinsic property of the 2p core-hole. Thus, core-excited states demonstrate an unprecedentedly fast purely electronic spin-flip dynamics, which is two orders of magnitude faster than that driven by nuclear motion in conventional spin-crossover~\cite{Auboeck2015}.

SF$_{\rm val}$ is a superposition of valence excited SOC states (see green bars in Fig.~\ref{fig:xas}b)). It turns out that it features a rather weak SOC, such that there is almost no dynamics happening within the considered time window of 15~fs. The inset in Fig.~\ref{fig:regimeI} shows population dynamics of different $M_S$ components of both quintet and triplet states without Auger decay. In fact one can see relatively slow redistribution of population between different quintet states and their $M_S$ microstates. Remarkably, the population of triplet states stays almost exactly zero during the considered time period. Therefore, this initial condition for valence excitation will not be discussed further.

\subsection{Regime II}
\label{sec:regimeII}
\begin{figure}
\centering
\includegraphics[width=1.0\textwidth]{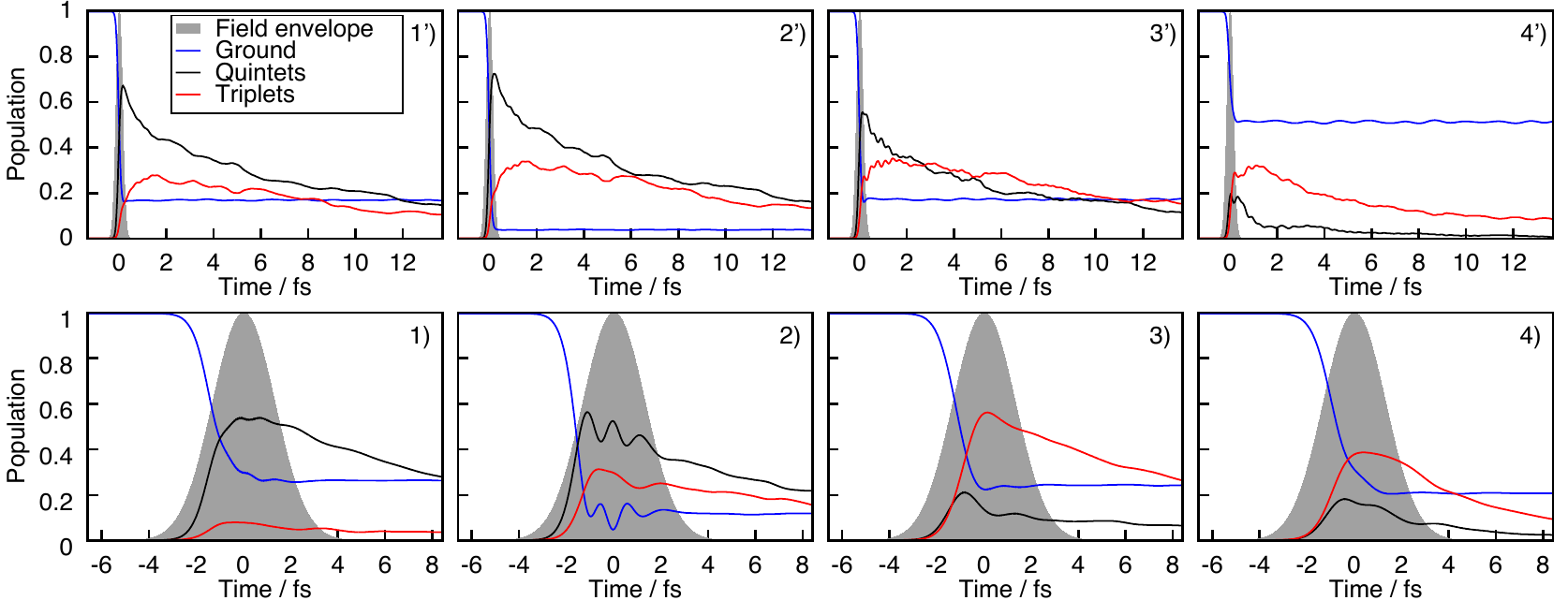}
\caption{Regime II: Spin dynamics initiated by the explicit field excitation with different carrier frequencies and bandwidths corresponding to: 1) $\hbar \Omega=$706.9~eV,  2) 708.4~eV, 3) 711.5~eV, and 4) 719.8~eV. Panels with primes correspond to bandwidth $\hbar/\sigma=5.0$\,eV, whereas without primes to $\hbar/\sigma=0.5$\,eV. See also arrows in Fig.\ref{fig:xas}a); for the spectral overlap with the SOC states, cf. Fig.~\ref{fig:xas}b). 
The field amplitude is $E_0=$2.5\,$\rm E_he^{-1}bohr^{-1}$ for pulses with 5.0\,eV bandwidth and $E_0=$1.5\,$\rm E_he^{-1}bohr^{-1}$ for 0.5\,eV.
Auger decay is accounted for.
Total populations of core-excited quintet and triplet states are shown by black and red lines, respectively. The envelope of the excitation pulse is shown as filled grey curve. The population of $M_S$-components of the ground and first two excited states is shown by the blue line.
\label{fig:regimeII}
}
\end{figure}

The spin dynamics upon excitation with ultrashort soft X-ray laser pulses polarized along $X$-direction (Fig.~\ref{fig:system}) with different carrier frequency and bandwidth/duration is shown in Fig.~\ref{fig:regimeII}. The respective energies and bandwidths in terms of the involved eigenstates are marked as numbered arrows and numbered ranges in Fig.~\ref{fig:xas}a) and b), respectively. Ranges with 0.5\,eV bandwith (8.3\,fs pulse duration) and 5.0\,eV bandwidth (0.8\,fs pulse duration) are denoted as intervals with and without primes, respectively. This notation is consistent with the names of panels in Fig.~\ref{fig:regimeII}. Here, the excitation frequencies correspond to  spectral regions with small and notable SOC mixing. 

For regime II, one can see similar ultrafast spin-flip dynamics as in regime I.  However, the population of all triplet states stays below 40\% within the time period of 15~fs in most cases.
As compared with regime I, most notable is the absence of the rapid oscillations. This is due to the fact that the temporal width of the pulse is longer than the 0.3~fs oscillation period dictated by SOC, i.e.\ the effect is smeared out. Further, compared to regime I, there are more slowly oscillating components in Fig.~\ref{fig:regimeII}. This can be traced to the fact that the initial state before excitation is  an incoherent thermal mixture of different $M_S$ components (Eq.~\eqref{eq:rho0}). Hence, the pattern of $\Delta M_S=0,\pm1$ transitions, which are possible upon excitation, changes.

The actual degree of quintet-triplet spin mixing is rather sensitive to the excitation conditions as follows from the comparison of different panels of Fig.~\ref{fig:regimeII}. 
Comparing these panels one notices similar oscillations, but noticeably different quintet/triplet ratios, reflecting the spin-mixing in the excitation range. In the case of smaller bandwidth, 1)-4), the oscillations are washed out almost completely. Further, the spectral selectivity with respect to the spin-mixing becomes even more pronounced. A slight modification of the excitation frequency from 706.9~eV to 711.5~eV, changes the quintet/triplet ratio at 15~fs from 0.1 to 7.7.

Note that the field strengths used here correspond to  intensities being of the order of $10^{17}$\,W/cm$^2$ and are barely accessible by current laser setups in the soft X-ray regime. 
However, these field strengths, despite of their large magnitudes, correspond to the weak field regime for soft X-ray wavelengths, with  Keldysh parameters $\gamma >  7$.  Moreover, the transition dipoles are quite small and the Rabi energy with respect to the strongest transition is $d_{\rm max}E_0=$ 2.7~eV and 1.6~eV for the broad and  for the narrow pulses, respectively. Therefore,  we do not  include strong field effects such as (multiphoton) ionization. In fact,  $E_0$ has been chosen merely to have an appreciable depletion of the ground state for illustration purposes. Indeed, the dynamics triggered by much weaker pulses ($10^{13}$\,W/cm$^2$
) qualitatively agrees with the strong ones~\cite{Wang_arxiv}. 



The variation of polarization directions leads to very similar dynamics even if $\vec{e}$ is aligned along $Z$ and $Y$ axes which correspond to the largest differences in the Fe--O bond lengths (Fig.~\ref{fig:system}). This means that for complexes with only slight distortions from octahedral symmetry the effect discussed here is of almost isotropic nature. However, for larger distortions we expect a stronger influence of polarization and thus, for the solution phase, averaging over polarization directions needs to be performed accounting for the free tumbling of the solute. For ordered phases (crystals) the variation of polarization might be an additional parameter to control the actual spin-state mixture and respective dynamics.

In order to validate the chosen setup we have also considered the inclusion of singlet states and of the RASPT2 correction. The singlet states (RASCI(0,1,2)) are not directly interacting with the initially prepared incoherent quintet spin-state mixture. The respective absorption spectra are shown in Fig.~\ref{fig:xas}a) and one can conclude that inclusion of singlets has a rather minor effect on the spectra. The same holds true for the respective dynamics where the populations of singlet states stay in most cases below 5\%, see Supplement. Thus, despite of the strong SOC, indirect coupling does not play a role in the considered time interval of 15\,fs.  

The same conclusion holds true for the effect of dynamic correlations introduced  via  RASPT2 corrections to the energies. This is already apparent from the absorption spectra in Fig.~\ref{fig:xas}a). Although the inclusion of this correction changes the effective couplings between the individual SF states due to changes of energy level separations, as a net effect the overall dynamics stays qualitatively almost the same, see Supplement. This underlines the good accuracy of the RASCI level of approximation at least for this system.
%
\subsection{Regime III}
\label{sec:regimeIII}

\begin{figure}
\centering
\includegraphics[width=1.0\textwidth]{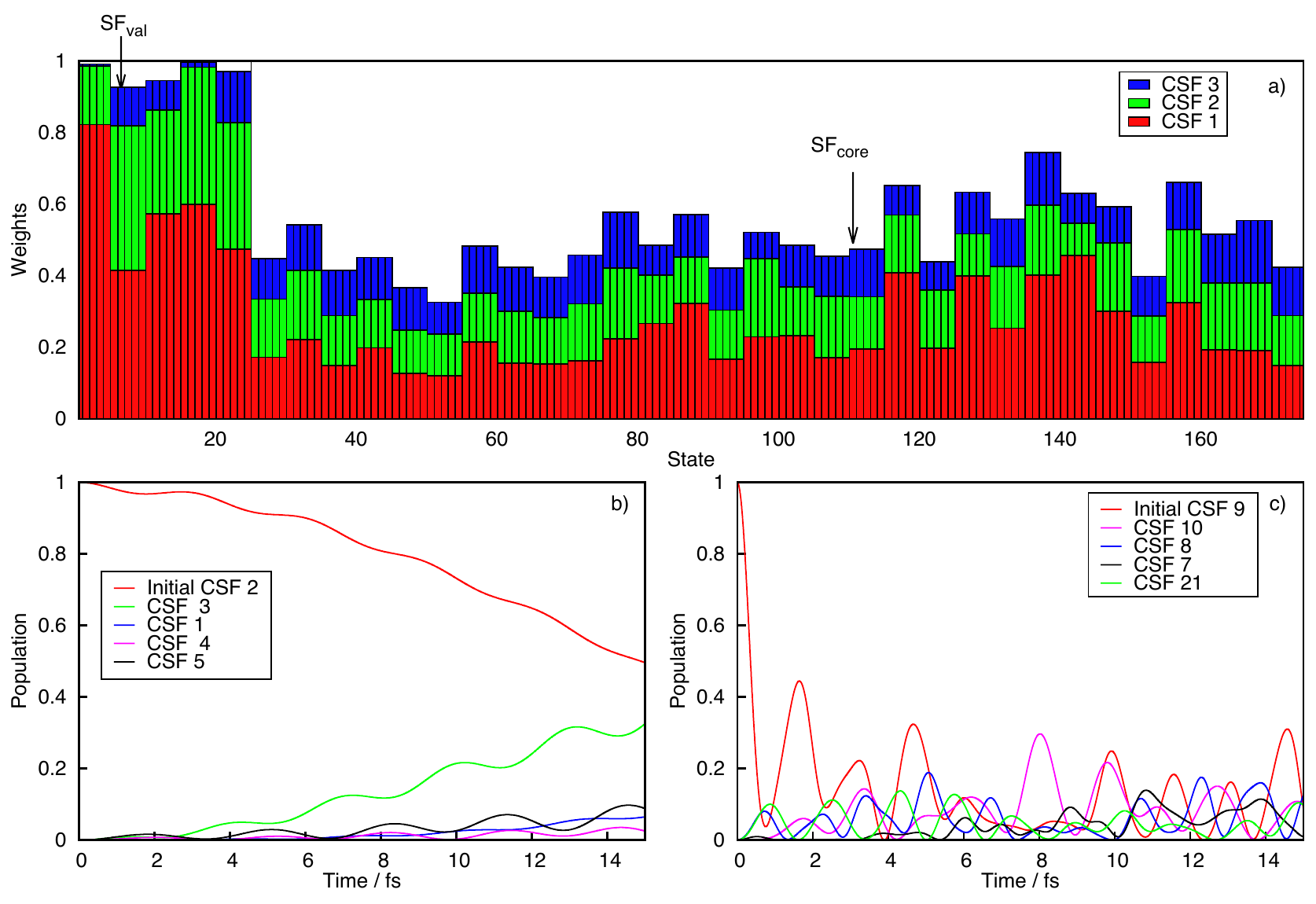}
\caption{Regime III: a) three largest CSF contributions to the CI wave function for the valence and core quintet SF states. The state number accounts for spin-degeneracy. b) and c) correspond to the propagation of the impulsively excited CSFs 2 and 9, respectively, representing the largest contributions to the spin-free states SF$_{\rm val}$ and SF$_{\rm core}$ marked with arrows in panel a).
\label{fig:regimeIII}
\label{Fig_corr}
}
\end{figure}

Electron correlation can be quite substantial for core excited states. 
To study its effect separate from the SOC, we have considered the dynamics of initially  prepared  electronic configurations (CSFs) belonging to the quintet spin-manifold. Fig.~\ref{Fig_corr}a) shows the three largest CSF contributions to the quintet CI wave functions accounting for spin-degeneracy. 
One can clearly see that for the core-excited SF states (6 -- 35) the weights of the dominant configurations are notably lower than for the valence ones (1 -- 5) being an indication of the stronger electron correlation.
 This fact is naturally reflected in the timescale and character of the dynamics.
 For illustration purposes we assume an instantaneous preparation of the two particular CSFs (2 and 9),  corresponding to the largest weights for the SF$_{\rm val}$ and SF$_{\rm core}$ discussed in Section~\ref{sec:regimeI}, respectively. Recall, that the latter have been chosen as representatives for the valence and core excited electronic states. 

The subsequent dynamics is shown in Fig.~\ref{Fig_corr}b) and c). In case of the CSF 2 (panel b)), some redistribution of the population occurs within 15~fs slightly modulated by the coherences with other electronic configurations. 
In contrast, the strong electron correlation in the core-excited state leads to an intricate dynamics with the ultrafast decay of the initial CSF 9 population with multiple revivals within the 15~fs time period (panel c)). 
The timescale of such dynamics is in general faster than that of the core-hole migration phenomena observed for ionized systems~\cite{Kuleff2014}. Thus, not only strong SOC itself but also electron correlation are important for the electron spin dynamics in the core-excited transition metal complexes.

\section{Conclusions and Outlook}
\label{sec:conclusions}
In this article, we have studied  ultrafast spin-flip dynamics driven solely by SOC, which should be typical for states having core-holes with a nonzero orbital momentum.  For this purpose we have formulated a density matrix based time-dependent restricted active space configuration interaction method suitable for the description of open quantum systems. This enabled to include phenomenological Auger decay. On the example of a prototypical third-period transition metal complex it was shown that soft X-ray light can trigger spin-flips, which are faster than the lifetime of the 2p core hole ($\approx$4~fs and $\approx$10~fs for Fe L$_2$ and L$_3$, respectively)~\cite{Ohno_2009}. Modifications of the pulse characteristics such as carrier frequency and pulse duration were shown to be effective in  controlling the actual spin mixture to quite some extent with modest effort.   

The notable dependence of spin state yields on the pulse parameters calls for an experimental verification. A possible  direct way to address such spin dynamics would be the upcoming time-resolved non-linear X-ray spectroscopy, e.g., stimulated resonant inelastic X-ray scattering (SRIXS)~\cite{Mukamel_2013, Haxton_2014, Zhang_2015}. In this technique the mixed-spin core states can be projected onto the manifold of pure-spin valence states, which are usually energetically well separated in transition metal complexes, see, e.g., discussion in Refs.~\cite{Bokarev2015, Golnak2016}. 
Thus, the relative SRIXS intensities in the respective energy ranges (0-1.5\,eV for quintets and 1.5-8.2\,eV for triplets in the case of \Fe system) would provide information on the time-evolution of the contribution of pure spin states to a mixed one.
Given the rapid progress in high harmonic generation~\cite{Dromey2006,Zhang2015,Popmintchev2015,Teichmann2016} and free-electron lasers~\cite{Picon_2016} and the expected establishment of time-resolved techniques such as SRIXS,~\cite{Mukamel_2013, Haxton_2014, Zhang_2015} the experimental proof of the effect discussed here and its use for manipulating spin dynamics appears to be within reach. 

\section*{Disclosure statement}
Authors declare no competing financial interests.

\section*{Funding}
This work was supported by the Deanship of Scientific Research (DSR), King Abdulaziz University, Jeddah under Grant No.\ D-003-435.


\end{document}